\documentclass[conference]{IEEEtran}
\usepackage{graphicx}
\usepackage[algoruled,boxed,lined]{algorithm2e}
\ifCLASSINFOpdf
\else
\fi

\begin{document}
%
\title{GANDALF: A fine-grained hardware-software co-design for preventing memory attacks}

\author{\IEEEauthorblockN{Gnanambikai Krishnakumar\IEEEauthorrefmark{1}, Patanjali SLPSK\IEEEauthorrefmark{2}, Prasanna Karthik Vairam\IEEEauthorrefmark{3} and  Chester Rebeiro \IEEEauthorrefmark{4}} \IEEEauthorblockA{Department of Computer Science and Engineering, IIT Madras\\ Email: \IEEEauthorrefmark{1}ambika@cse.iitm.ac.in, \IEEEauthorrefmark{2}slpskp@cse.iitm.ac.in, \IEEEauthorrefmark{3}pkarthik@cse.iitm.ac.in, \IEEEauthorrefmark{4}chester@cse.iitm.ac.in}}


%



\def\gandalf {{\sf Gandalf}}

\maketitle

\begin{abstract}
Reading or writing outside the bounds of a buffer is a serious security vulnerability that has been exploited in numerous occasions. These attacks can be prevented by ensuring that every buffer is only accessed within its specified bounds. In this paper we present \gandalf{}, a compiler assisted hardware extension for the OpenRISC processor that thwarts all forms of memory based attacks including buffer overflows and overreads.The feature associates lightweight base and bound capabilities to all pointer variables, which are checked at run time by the hardware. \gandalf{} is transparent to the user and does not require significant OS modifications. Moreover it achieves locality, thus resulting in small performance penalties. 
\end{abstract}


%
\IEEEpeerreviewmaketitle

\section{Introduction}
Even though twenty years have gone by since Aleph One's benchmark paper in~\cite{one:96}, buffer overflows remain one of the most exploited vulnerability in C and C++ programs. Malware that overflow buffers corrupt system or other critical memory regions forcing execution of an attacker specified code. This has lead to privilege escalation attacks, malfunction of applications, network penetration, or denial of service. More recently a variant of buffer overflow, called the {\em Heartbleed bug}\footnote{http://heartbleed.com/}, overreads buffers allowing an attacker to read sensitive information from the program space. This has been used, for instance, to leak information such as certificates, digital signatures, and passwords from a web server that uses the OpenSSL library~\footnote{https://www.openssl.org/}.

Over the years several preventive measures have been introduced into systems. One of the first, was the use of smart pointers in programming languages like Java and C\# to perform bound checking on buffers. While these languages stymie buffer overflows, they unfortunately cannot replace the huge amounts of C and C++ codes that is in use and cannot be applied for system programming. Modifications were also made to the compiler, for example by the use of canaries and {\em address space layout randomization} (ASLR), which made it more difficult to exploit buffer overflows but did not completely eliminate them. The weakness of these approaches were that they did not try to stop buffers from overflowing but rather tried to prevent overflowing buffers from causing damage. In due course, attackers found ways to execute their payloads in the system in-spite of canaries and ASLR. 

The wide spread use of buffer overflows in several malware, worms, and viruses, dictated prevention mechanisms to be implemented in the hardware. Intel was one of the first to introduce such a mechanism with the NX bit in their processors. This prevented attack payloads executing from the stack and other data segments in the program. Attackers soon found ways to bypass this with attacks such as the return-to-libc and the return-oriented-programming (ROP) attack~\cite{roemer:12}. More recently Intel introduced the MPX instruction extensions~\footnote{https://software.intel.com/en-us/articles/introduction-to-intel-memory-protection-extensions} to further fortify their processors against ROP attacks. Other processor vendors are following suit. For example, ARM introduced authenticated pointers~\footnote{https://community.arm.com/groups/processors/blog/2016/10/27/armv8-a-architecture-2016-additions} in their recent additions. 

In the research community there have been several works to prevent buffer overflows by modifying the processor architecture. The approaches can be broadly classified as follows : the use of a shadow stack (such as~\cite{kuperman:06,lee:03}), tagged memories (such as~\cite{song:16}), xor encrypted function calls (such as~\cite{xiao:06}), signed memories (such as~\cite{lie:03}), and by control flow integrity checks (such as~\cite{davi:15}). Encrypted function calls, tagged, and signed memory techniques require programming acumen to differentiate between sensitive and non-sensitive parts of the code. Thus can be subject to human error leading to exploits. On the other hand shadow stacks are limited to preventing exploits based on corrupting return addresses. Other vulnerabilities such as corruption of function pointers, modification of local variables, and buffer overreads cannot not be prevented. Further, implementations of shadow stack require memory that is typically isolated from the program. This could lead to execution overheads due to loss in locality.

In this paper we propose \gandalf{}; a  mitigation technique for buffer overflows and overreads. The motivation is to make security transparent to programmers thus preventing human error. Our proposal  is (almost) operating system agnostic and only requires architecture and compiler changes. Furthmore, \gandalf{} provides fine-grained security, which is capable of protecting each pointer individually and blocks all known memory corruption attacks in the stack and heap. It also prevents attacks like Heartbleed which overreads buffers.

\begin{figure}
    \centering
    \includegraphics[scale=0.45]{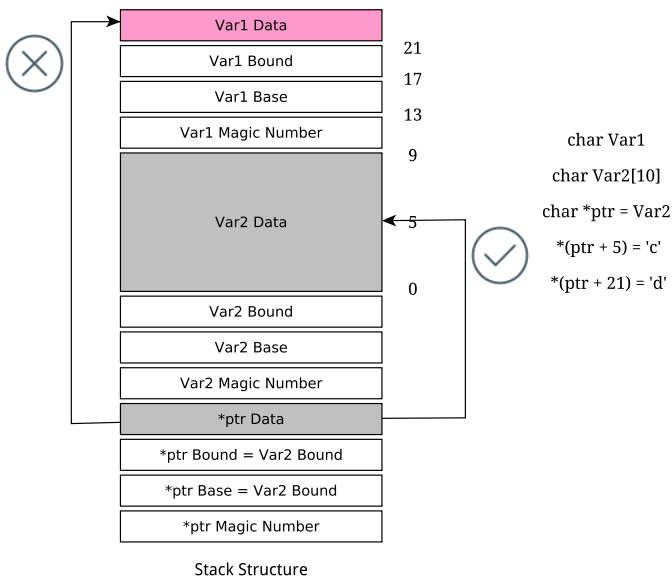}
    \caption{Stack Structure with \gandalf{} enabled. The pointer {\tt ptr} can only access the buffer {\tt var2} because its capabilities are set by the limits of {\tt var2}. Any attempt to read or write to a location beyond {\tt var2} will result in an exception.}
    \label{fig:stackat}
\end{figure}

\gandalf{} works by pre-pending every declared pointer in the program with three capabilities: base, bound, and a magic number as shown in Figure~\ref{fig:stackat}. The base and bound define the capabilities of the pointer, while the magic number allows quickly to identify the metadata. It helps prevent unauthorized changes to the base and bound values. Each load and store, is first authenticated by the pointer's capabilities and only then allowed to complete. If a load or a store fails this test, then an illegal instruction exception is sent to the program terminating it. 

We implemented \gandalf{} in an OpenRISC environment and tested it with different memory corruption based exploits. It required about 500 lines of Verilog code to be changed / added/modified in the processor.

The rest of this paper is organized as follows: Section~\ref{overview} describes the overall sketch of our solution. Implementation of the solution is detail is covered in Section~\ref{implementation} in terms of changes made to the compiler, operating system and the hardware. Some security arguments supporting our scheme is presented in Section~\ref{securityarg}. The Overhead of our solution and some optimizations to improve the performance are discussed in Section~\ref{overheads}. Finally, we conclude our findings in Section~\ref{conclusion}.

\section{GANDALF: A least effort memory defense scheme}
\label{overview}

\begin{figure*}
    \centering
    \includegraphics[scale=0.30]{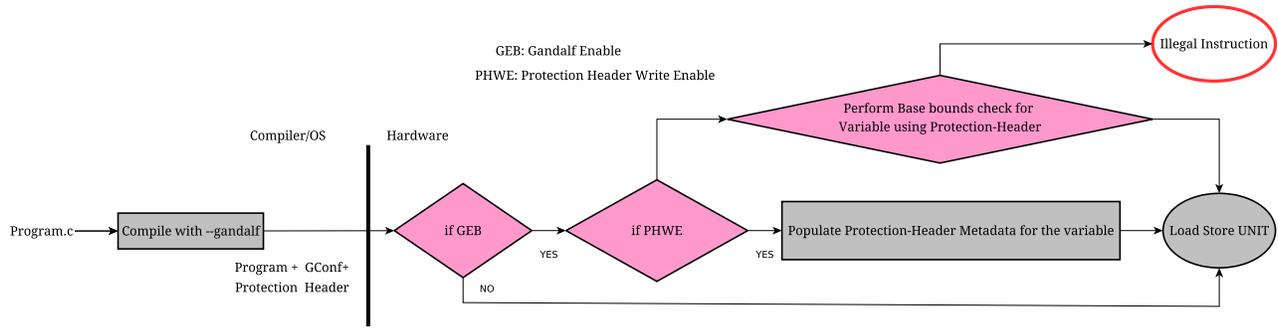}
    \caption{Flowchart describing \gandalf{} scheme Grey indicates the steps done at compiler level and Pink indicates the steps done by the hardware}
    \label{fig:gandalf}
\end{figure*}

In this section we describe the working of \gandalf{}, the flowchart is shown in Figure~\ref{fig:gandalf}. \gandalf{} performs memory out of bound access checks for every named variable in the user program. Every variable in the user program is augmented with metadata, called the {\em Protection-Header} that specifies its legal boundaries. The Protection-Header is then used by the hardware to perform base and bounds check during every memory access. When a variable tries to write or read to a memory location outside its bounds, \gandalf{} triggers an exception causing the program to terminate.

\gandalf{} ensures that the programmer is oblivious to the presence of the program-header as well as the bound checks made by the hardware. The Protection-Header and the \gandalf{} hardware configuration (Protection-header) details are added by the instrumented compiler when the user code is compiled. The Protection-header is added to instruct the processor to switch between \gandalf{} and the normal mode of operation. Enabling \gandalf{} involves {\bf (a)} populating the Protection-Headers and {\bf (b)} using the Protection-Headers for bounds check. When the program starts, \gandalf{} uses Protection-Header to instruct the hardware to populate the Protection-Headers. Once the headers are up to date, it instructs the hardware to enable bound checking. Protection-Headers comprise of the base, bound and the Magic number for every variable used in the program, as shown in Figure~\ref{fig:stackat}. 

Data in the program can be of four types, namely a) scalar variables, b) arrays, c) pointers and d) system variables (such as return pointers and frame pointers). The code to populate Protection-Headers for cases a) and b) is introduced by the compiler after analyzing the variable tyes. However, in case of pointers, the compiler ensures that they inherit the base and bound values based on the variable it points to at runtime. 

All load and store instructions from the user program are routed through the \gandalf{} hardware since these are the only instructions that manipulate memory in any load-store architecture. \gandalf{} hardware performs bound check based on the metadata available in the Protection-Header corresponding to the address used in the load or store. Firstly, the load-store instruction is decoded to identify the program variable associated with the instruction. Next, we index into the Protection-Header to extract the base and bound associated with the variable. Finally, we check if the indexed value is within the valid limits specified by the base and bound. 

Programmers can choose to enable or disable \gandalf{} for every program using a compiler option. The compiler instructs the \gandalf{} hardware to perform checks using the Protection-Headers. Therefore, Linux and user programs can be instrumented very easily to use the Gandalf hardware for security. Additionally, programs (existing programs inclusive) that do not have Protection-Header will bypass the Gandalf hardware logic and execute normally.

\subsection{Assumptions}
We make the following assumptions about the attacker's capabilities, hardware and the software.

\begin{enumerate}

\item  The stack structure is intuitive to attackers. Specifically, the location and contents of Protection-Headers and data are not assumed to be oblivious to attacker.

\item  The attacker cannot modify the executable binary of victim program. However, the attacker can provide run time inputs (payloads) to the program, in order to modify the control flow.
\item  Since the data accesses are made only using the register-indirect addressing mode in OpenRISC, every memory access is of the form \textit{load rB, rA(I)} or \textit{store rA(I), rB}. For array and pointer based accesses, we expect the compiler to use the array's base address in (rA). For non-array variables, we require the compiler to use the register r2 (which stores the stack frame pointer) as the base. It is to be noted that the gcc compiler already does this.

\item  The two bits in SPR registers, namely Gandalf Enable bit (GEB) and the Protection Header write enable bit (PHWE), are to be saved and restored during context switches. We have made necessary modifications to the Linux source code to achieve the same.

\item We have not tested attacks that overflow the data/system variables present on the heap, though our scheme can be directly or with slight modifications used to handle this.

\end{enumerate}

\section{Implementation}
\label{implementation}
\gandalf{} requires modifications to be made in the hardware and compiler. Typically no modifications are required in the OS. However in the current version of the Linux kernel (ver 4.4.0-de0\_nano) that was used, minor modifications were needed in the kernel to save context. This section provides a top-to-down overview of the \gandalf{} implementation on OpenRISC.

\subsection{Compiler Changes}
Every program compiled for the system has an option to turn on/off \gandalf{} independent of the other programs. This is done by a compile time switch that respectively enables/disables \gandalf{}.
When enabled, the compiler programs the following additional steps into the executable image.

\begin{itemize}
    \item When the program starts it enables \gandalf{} in the hardware. In OpenRISC this is implemented by setting the $17^{th}$ bit in the special purpose register (the SPR). This bit is referred to as the \gandalf{} enable bit (GEB). 
    \item For every named variable the protection-header data fields including the magic number, base, and bound are populated.
    While the base and bound are obtained from the program flow, the magic number is filled with its own address as follows. \\
    \hspace{7mm}${\tt store [Magic\_addr], magic\_addr}$, 
    where [$\cdot$] denotes the value stored in the address.
    For example, if 0x80012340 is the magic address then the magic number is 0x80012340. This simple step allows {\bf (a)} unique magic numbers for every live protection-header in the program and {\bf (b)} an easy way to test the magic number. To perform the test, we simply need to check if the loaded data is equal to the corresponding address. 
    
    To write to protection-headers a special store instruction is used.
    The processor is made to distinguish between a normal store and this special store by setting the $18^{th}$ bit in the SPR. This bit is called the protection-header write enable bit (PHWE). 
    
    \item Resetting the ${18^{th}}$ bit in the SPR indicates that all subsequent loads and stores have to be validated for memory overflows or overreads. This validation happens in hardware.
\end{itemize}
    
\subsection{Hardware Changes}
    When the \gandalf{} enable flag (GEB) is set and the protection-header write enable bit (PHWE) is 0, hardware extensions to OpenRISC validates every load and store as shown in Algorithm 1. 
    \begin{itemize}
        \item If GEB is set and  PHWE is 0, the execute stage, generates three addresses in addition to the effective address of the load / store. These three addresses correspond to the magic number, base pointer and bound pointer for the corresponding effective address.
        \item In OpenRISC's load-store unit, data is fetched as follows.
            \begin{enumerate}
                \item Fetch data from address corresponding to the magic number and do the following check to ensure that it is indeed the magic number.
                \begin{equation}
                [Magic address]== Magic address  \end{equation},
                where [$\cdot$] denotes the value stored in the address.
                \item If the magic data sanity check passes then the data from address stored in base pointer is fetched  and the effective address is then compared with it. If the effective address is greater than the base then we fetch the bound address
                If the effective address is less than the bounds then the corresponding load/store instruction is performed. Any violation will trigger an exception which will cause the program to halt.
                
            \end{enumerate}

\end{itemize}         
      
\vspace{5mm}
\begin{algorithm}
\SetAlgoLined
\If{ $ GEB~\&\&~!PHWE $ }{
	\textbf{Set} mismatch = true\;
	\uIf{ $ [MagicAddress]== MagicAddress $ }{
		\If{ $ BaseAddress < EffectiveAddress $ }{
			\If{ $ BoundAddress > EffectiveAddress $ }{
				mismatch = false\;
			}		
			
		}

	}
	
	\If{ $ mismatch == true $ }{
		raise $ mismatch  exception $\;
	}
	\Else{
		execute instruction\;
	}
}
\ElseIf{$ GEB~\&\&~PHWE $ }{
 \tcc{Populate protection-header\hspace{-0.2cm}}
	set mismatch = false\;
	execute instruction\;
}
\Else{
\tcc{Normal loads and stores\hspace{-0.4cm}}
  \textbf{Set} mismatch = false\;
  execute instruction\; 
}     

\caption{Hardware Protection Checks}
\end{algorithm}

\subsection{Linux Kernel Changes}
 Since the protection scheme is selectively enabled or disabled from process to process, the operating system has to save and restore the GEB and PHWE bit for every process. The ${\tt thread\_start}$ and ${\tt thread\_switch}$ functions in the Linux kernel to save the state of these bits on a per-process basis. It may be noted that if these bits were implemented in the flags register, whose state is already saved on a per-process basis, then the OS would require no alteration.

\section{Security Argument}
\label{securityarg}
 In order for memory corruption to occur, the attacker must be able to manipulate the effective address to point to the target data as shown in Figure~\ref{fig:stackat}. This can only be possible using one of the following methods:
\begin{itemize}
    \item \textbf{Overflowing a memory segment}\\
           \textbf{Claim:} This is not possible because each named variable is tagged with a protection-header and the attacker cannot read and write into memory locations that are outside the bounds.\\
    \item \textbf{Manipulating the content of the protection-header}\\
           \textbf{Claim1:} A write into the protection-header memory segment is possible only when the $18^{th}$ bit in the Special purpose register is set and since the user has no control over this bit, any write into the protection-header memory segment by the attacker will result in a exception.\\
           \textbf{Claim2:} A write into the protection-header memory segment using a normal store will trigger the \gandalf{} check routine in hardware and since the protection-header memory segment will not have its appropriate protection-header data, an exception is triggered.\\
    \item \textbf{Generating a spurious protection-header}\\
           \textbf{Claim1:}The user cannot setup the protection-header content via the payload for the following reason: Once the spurious-protection header content is set, the attacker can perform malicious memory access if he can generate effective addresses that are within the range of malicious base and malicious bound. OpenRISC uses a relative displacement addressing scheme for accessing data and hence the attacker needs to manipulate either the base or the offset part of the address. The attacker has no control over the base and if he tries to generate a spurious effective address by manipulating the offset, the protection-header of the corresponding base address will detect it.  \\
           For example,
                if the attacker tries to do
                    int a[15];
                    int *p =\&a[14]+4;
                    p will inherit a's protection-header content and hence any subsequent access via p will be treated as illegal.
\end{itemize}
The above claims show that our scheme protects memory segments from getting corrupted by including the protection-header.

\section{Overheads and Optimizations}
\label{overheads}
    \gandalf{} offers additional security at the cost of performance and memory that is required to execute the programs. We can divide \gandalf{}\'s operations into a) populating the Protection-Headers and b) performing base and bounds check at run-time. 
    
\subsection{Protection-Header Population}
    
    Populating the Protection-Headers requires additional store instructions per program variable, in order to populate the base, bound and the magic number. The Protection-Header population for scalar, array and system variables happens in the prologue of every function. For pointers, the protection-header is added whenever the pointer is initialized. This operation increases the overall time required for the program to execute. Additionally, the memory overhead required for storing the Program-Header is also high. \\
    
    \textbf{Optimization:} OpenRISC uses a store buffer to do lazy writes and hence the penalty for writing data into memory will not be felt by the processor. \\
    
\subsection{Base Bound Checks}
    Performing base and bounds check at run-time requires additional ALU operations to compute 3 additional effective addresses (for base, bound and the magic number). More importantly, 3 extra loads need to be performed to retrieve the base, bound and the magic numbers. The extra loads are issued for both load and store instructions making the clock cycles required per operation very high.\\
    
    \textbf{Optimization:} Since the protection-header data has spatial locality with respect to the current data segment, it is mostly likely to be present in the L1 cache. Thus an increase in the L1 cache size may have significant benefits. A further optimization could be to maintain 3 on chip registers to store the protection-header data.\\

\subsection{Compiler Overheads}
Adding additional instructions will cause the program to bloat.\\

    \textbf{Optimization:} The increase in size is less than $30\%$  
    
\section{Conclusion}
\label{conclusion}
Memory corruption attacks are a serious threat to software. Over the last 20 years, several works have been published in this regard, however recent design trends have necessitated the need for a robust hardware model which requires less design overhaul. In this work, we introduce \gandalf{}, a light-weight and robust hardware-software ecosystem which prevents such attacks. We show that the scheme achieves fine grained security  through our security argument. Further,\gandalf{}  nips buffer overflows at the bud, it prevents all forms of attacks including the format string vulnerabilities, ROP attacks, and return-to-libc attack.
The hardware or software changes is minimal when compared to the other schemes that currently exist in practice. 


\section*{Acknowledgment}
We would like to thank Arjun Menon for his inputs during the initial stages of discussion. We would also like to thank Prof. Kamakoti and ISEA for sponsoring our travel and our lab mates for their endless support. We would like to extend our thanks to the organizing committee of CSAW for their timely support.

\bibliographystyle{IEEEtran}
\bibliography{bare_conf}
%

\end{document}